\documentclass[a4paper,12pt]{article}
\usepackage{amssymb}
\usepackage{latexsym}
\usepackage{amsfonts}
\input epsf
\newcommand{\be}{\begin{enumerate}}

\newcommand{\ee}{\end{enumerate}}
\newcommand{\E}{\mathbb E}

\newcommand{\ind}{{\mathrm{1\! \! \hspace{0.03 cm} l}}}

\newcommand{\noi}{\noindent}

\newcommand{\qal}{q_{1-p}}
\newcommand{\Real}{\mathbb R}
\newcommand{\RPLus}{\Real^{+}}

\newcounter{numero}

\begin{document}
\baselineskip0.9cm
\title{{\bf  Quasi-conjugate Bayes estimates for GPD parameters
and application to heavy tails modelling}}
\author{
{\bf \large Jean Diebolt  \hspace{3 cm} Mhamed-Ali El-Aroui}\\
{\small \hspace{-2.5 cm} CNRS, Universit\'e Marne-la-Vall\'ee \hspace{3.5 cm} ISG de Tunis} \\
{\small 5 Bd Descartes. 77454 Marne-La-Vall\'ee. France.  \hspace{0.5 cm} 41 Av. de la Libert\'e, Bardo 2000. Tunisia.} \\
{\tt \small diebolt@math.univ-mlv.fr \hspace{2 cm} Mhamed.Elaroui@isg.rnu.tn}\\
\\
{\bf \large Myriam Garrido \hspace{3 cm} St\'ephane Girard}\\
{\small \hspace*{1 cm} Dept. MI, ENAC \hspace{3.1 cm} SMS-LMC, Universit\'e Grenoble I}\\
\hspace*{-1 cm} {\small Av. E. Belin. 31055 Toulouse. France. \hspace{1.5 cm} BP 53, 38041 Grenoble. France. } \\
{\tt \small garrido@recherche.enac.fr \hspace{2.1 cm} Stephane.Girard@imag.fr}}
\date{}
\maketitle
\noindent
{\bf Abstract --} We present a quasi-conjugate Bayes approach for estimating Generalized Pareto Distribution (GPD) parameters, distribution tails and extreme quantiles within the Peaks-Over-Threshold framework. Damsleth conjugate Bayes structure on Gamma distributions is transfered to GPD. 
Posterior estimates are then computed by Gibbs samplers with Hastings-Metropolis steps. 
Accurate Bayes credibility intervals are also defined, they  provide assessment of the quality of the extreme events estimates. 
An empirical Bayesian method is
used in this work, 
but
 the suggested approach could incorporate
prior information.
It is shown that the obtained  quasi-conjugate Bayes
estimators compare well with the GPD standard estimators when simulated and real data sets are studied. 

\noindent
{\bf Key words --} Extreme quantiles, Gamcon~II distribution, Generalized Pareto Distribution,
Gibbs Sampler,  Peaks over thresholds (POT).

\section{Introduction}
Motivated by univariate tail and extreme quantile estimation,
our goal is to develop new Bayesian procedures for making statistical
inference on the shape and scale parameters of
Generalized Pareto Distributions (GPD) when used to model heavy tails
and estimate extreme quantiles.

Let us assume that observations of
a studied phenomenon $x_1, x_2, \dots, x_n$ are issued from
independent and identically distributed (i.i.d.)
random variables $X_1, X_2, \dots, X_n$ with unknown common
distribution function (d.f.)~$F$. 
Suppose that one needs to estimate either extreme quantiles $\qal$
of~$F$ (i.e., $1 - F(\qal) = p$ with $p$ $\in$ $(0, \, 1 / n]$
typically), or the extreme tail of~$F$ (i.e., $1 - F(x)$ for $x
\ge x_{n, \, n}$, where $x_{1, \, n} \le \dots \le x_{n, \, n}$
denote the ordered observations). It is usually recommended to use
the Peaks Over Threshold (POT) method (described for
example in Davison and Smith (1990) or in the monographs
Embrechts {\em et al.} (1997) and Reiss and Thomas
(2001)), where only observations $x_i$ exceeding a
sufficiently high threshold~$u_n$ are considered.
In view of the theorem of Balkema and de Haan (1974) and Pickands (1975)
the probability
distribution of the $k$ $=$ $k_{n}$ positive excesses $y_{j}$ $=$
$x_{n-j+1, \, n} - u_n$ for $j = 1, \dots, k$, where $x_{n-k, \, n}$
$<$  $u_n$ $\le$ $x_{n-k+1, \, n}$, can be approximated for
large~$u_n$ by a GPD$(\gamma, \, \sigma)$ distribution with scale
parameter $\sigma
>0$ and shape parameter~$\gamma$. The d.f.~of GPD$(\gamma, \,
\sigma)$ is
\begin{eqnarray}
F_{\gamma, \, \sigma}(y) \, = \,
\left\{
\begin{array}{ll}
\displaystyle
1 \; - \; \left( 1 \; +  \;
\frac{\gamma y}{\sigma} \right)_{+}^{-1/\gamma}
& \mbox{if} \; \gamma \neq
0 \\
\\
\displaystyle
1 \; - \; \exp{\left( - \frac{y}{\sigma}
\right)}
& \mbox{if} \; \gamma = 0 \, ,
\end{array}
\right.
\end{eqnarray}\label{equ1}
with $y_{+} = \max(y, \, 0)$, where $y$ $\in$ $\RPLus$ when $\gamma \ge 0$,
and $y$ $\in$ $[0, \, -\sigma/ \gamma]$ when $\gamma < 0$.

The shape and scale parameters of the approximating GPD are estimated
on the basis of the excesses above $u_n$. The estimates are then
usually plugged into the GPD d.f.~and extreme quantile estimates
are deduced. In this perspective, good estimation procedures for
the shape and scale parameters of a GPD on the basis of
approximately i.i.d.  observations are necessary for accurate tail estimation. 
Estimating the shape  and scale parameters,~$\gamma$ and~$\sigma$,
is not easy. 
Smith (1987) has shown that estimating GPD parameters
by maximum likelihood (ML) is a non regular problem for $\gamma < -1/2$.
Besides, the ML estimators 
may be numerically hardly tractable, see Davison and Smith (1990)
and Grimshaw (1993). 
Moreover, the properties of MLE's meet their asymptotic
theory only when the sample size (in our case the number~$k$ of
exceedances) is larger than about~$500$.
Many alternative estimators have been proposed: Hosking and Wallis (1987)
linear estimators are based on probability weighted moments (PWM);
they are easily computed and reasonably
efficient when $-0.4 < \gamma < 0.4$
approximately  (see
also the comparative study of Singh and Guo, 1997). \nocite{Singh97}
Castillo and Hadi (1997) \nocite{Castillo97} have
proposed other estimators based on the elemental percentile method (EPM),
involving intensive computations. Their numerical simulations show that
EPM estimators are more efficient than PWM ones only when $\gamma < 0$.


Semiparametric estimators of~$\gamma$ along with
related estimators of extreme quantiles have been intensively
studied. For example,
the Hill estimator presented by Hill (1975) and studied in
Haeusler and Teugels (1985) and Beirlant and  Teugels (1989),
among many others.
Two classic extensions of the Hill estimator are:
The moment tail index estimator (denoted hereafter MTI(DEdH)) of Dekkers, Einmahl and de Haan (1989);
The Zipf estimator, see Schultze and Steinebach (1996), and its
generalization by Beirlant, Dierckx and Guillou (2001), denoted
hereafter ZipfG.
Most of these semiparametric estimators  do not perform much better
than  parametric ones when applied to sets of excesses.
Only the ZipfG estimator seems to outperform the other ones.

In this paper, a new Bayesian inference approach
for GPD's with $\gamma > 0$ is introduced.
 In a number of application areas such as structural
reliability (see for example Grimshaw, 1993) and excess-of-loss reinsurance
(see Reiss and Thomas, 2001), tail estimation based on small
or moderate data sets is needed. 
In such situations Bayes procedures can be used to
capture and take into account all available information
including expert information even
when it is loose. Moreover, in the realm of tail inference,
evaluating the imprecision of estimates is of vital importance.
Bootstrap methods have been suggested to assess this imprecision. But
standard bootstrap based on larger values of ordered samples
is known to be inconsistent, whereas standard bootstrap based on excess samples
has not second-order coverage accuracy and is imprecise when sample sizes
are not
extremely large (e.g., Bacro and Brito (1993), 
Caers, Beirlant and Vynckier, 1998).
On the contrary, in the Bayesian context, credibility regions
and marginal credibility intervals for GPD parameters and related high
quantiles
provide a non-asymptotic geometry of uncertainty directly based on outputs
of the procedure, thus shortcutting bootstrap.
For all these reasons (expert information, credibility intervals)
easily implementable Bayesian inference
procedures for GPD's are highly desirable to study excess samples in the
scope of POT~methodology.

The restriction $\gamma > 0$ is not too damaging,
since several major application areas are connected to
heavy-tailed distributions. Our approach can also be tried for
data issued from distributions suspected to lie in Gumbel's
maximum domain of attraction (DA(Gumbel)) where $\gamma= 0$. 
In the latter case, direct Bayesian analysis of the exponential
distribution can be made in parallel (see Appendix A).

For other papers on Bayesian approaches to high quantile
estimation, see, e.g., Coles and Tawn (1996), Coles and Powell (1996),
Reiss and Thomas (1999), 
Tancredi {\em et al.} (2002),
Bottolo {\em et al.} (2003),
and the monographs Reiss and Thomas
(2001) and Coles (2001) along with references therein.

Our starting point is a representation of heavy-tailed GPD's
as mixtures of exponential distributions with a gamma mixing distribution.
Since the Bayesian conjugate class for gamma distributions is documented
(Damsleth, 1975) we only have to transfer it to GPD's.
As described in Section~2, this provides a natural Bayesian quasi-conjugate
class for heavy-tailed GPD's.
Even though Bayes estimators have no analytical expressions, the quasi-conjugate structure
makes the Bayes computations very simple, the convergence of the MCMC algorithms very quick
 and gave a high parsimony to the global approach with an intuitive interpretation of the hyperparameters.

Section~3 compares our Bayes estimates to
ML, PWM, moment tail index MTI(DEdH)
and generalized Zipf (ZipfG) estimates on excess samples
through Monte Carlo simulations. Section~4 is devoted to
benchmark real data sets. Finally, Section~5 lists some conclusions and presents forthcoming research projects.


\section{Bayesian inference for GPD parameters}
The standard parameterization of heavy-tailed GPD distributions
described by (\ref{equ1}) when $\gamma >0$ is now replaced
by a more convenient one depending on the two positive parameters
$\alpha$ $=$ $1/\gamma$ and $\beta$ $=$ $\sigma/\gamma$.
The re-parameterized version GPD$(\alpha, \, \beta)$ has 
probability density function (p.d.f.)
\begin{eqnarray}
f_{\alpha, \, \beta}(y) \, = \,f \left( y \left\vert \, \alpha, \, \beta \right. \right) \, = \,
\frac{\alpha}{\beta} \,
\left( 1 \; + \; \frac{y}{\beta} \right)^{- \alpha \, - \, 1} ,
\quad y \ge 0 \, .
\label{eq:GPDdensite}
\end{eqnarray}
We assume in the following that we have observations
${\mathbf y}$ $=$ $(y_1, \dots, y_k)$
which are realizations of i.i.d.~random variables $Y_1, \dots, Y_k$
approximately issued from~(\ref{eq:GPDdensite}).
Typically, they represent excesses above some sufficiently high
threshold~$u$.  The latter means that for each $j \leq k$,
there exists an integer $i\leq n$ such that
$Y_{j}$ $=$ $X_{i} - u$, $X_{i} > u$,
where the $X_{i}'s$ are assumed i.i.d.~and issued
from a distribution in Fr\'echet's maximum domain of attraction:
DA(Fr\'echet). 
Remark that the case where the common distribution of the
$X_{i}'s$ is in DA(Gumbel)
can also be covered by considering the limiting situation
$\alpha \to +\infty$ and $\beta \to +\infty$ with $\beta/\alpha
\to \sigma > 0$ (see Appendix~A).

\noi Our starting point is the following mixture representation
for~(\ref{eq:GPDdensite}),
see Reiss and Thomas (2001) page~157:
\begin{eqnarray}
f \left( y \left\vert \,
\alpha, \, \beta \right. \right) \, = \,\int_0^{\infty}
z \, e^{- y z} \,
g \left( z \left\vert \, \alpha, \, \beta \right. \right) \, dz
\, ,
\label{eq:melange}
\end{eqnarray}
where for $z \geq 0$, $g ( z \vert \, \alpha, \, \beta )$  $=$
$[ \beta^{\alpha} / \Gamma ( \alpha ) ] \, z^{\alpha  -1}
e^{- \beta z}$ is the density of the
Gamma$(\alpha, \, \beta)$ distribution with shape
and scale parameters~$\alpha$ and~$\beta$. The previous representation stands only in DA(Fr\'echet) as $\alpha$ and $\beta$ have to be non-negative (as parameters of a gamma distribution) which implies $\gamma>0$.

There is no Bayesian conjugate class for GPD's.
Nevertheless, as shown below, the mixture form~(\ref{eq:melange}) allows us
to make use of the conjugate class for gamma distributions
to construct a suitable quasi-conjugate class for GPD's.
\subsection{Transferring conjugate structure from Gamma to GPD}
\label{conjugate}
According to Damsleth (1975), the description of the
conjugate class for gamma distributions
relies on the so-called type~II Gamcon distributions: for $x>0$, the
density of the Gamcon~II($c, \, d$) distribution
with parameters $c>1$ and $d>0$ is
\begin{eqnarray}
\xi_{c, \, d } \left(x \right) \, = \, I_{c, \, d}^{-1} \;
\Gamma(d x \, + \, 1) \left(\Gamma(x)\right)^{- d}
(c \, d)^{- d x} \,  ,
\label{eq:gammaconjII}
\end{eqnarray}
where
$I_{c, \, d} = \int_{0}^{\infty}
\Gamma(dx \, + \, 1)(\Gamma(x))^{-d} (c \, d)^{- d x} \, dx$.
Let in the following  ${\mathbf z}$ $=$ $(z_{1}, \dots,z_{k})$ denote a $k$-sample
of realizations of i.i.d.~random variables $Z_1, \dots, Z_k$
issued from Gamma$(\alpha, \, \beta)$. According to Damsleth (1975),
Theorem 2, the conjugate prior density 
on $(\alpha, \, \beta)$ with hyperparameters $\delta > 0$ and
$\eta > \mu > 0$ is given by
\hspace{0.3 cm} $\pi_{\delta, \, \eta, \mu} (\alpha, \, \beta) \,  = \,
\pi (\alpha)  \, \pi \left(\beta \left\vert\, \alpha \right. \right) \,$ \hspace{0.3 cm}
where
$\pi(\alpha)$ is the density of
Gamcon~II($c = \eta / \mu , \, d = \delta$) and
$\pi(\beta \vert \, \alpha)$ is the density of
Gamma($\delta \alpha \, + \, 1 , \, \delta \eta$).
Then, the conditional density of~$\alpha$ given~${\mathbf z}$, denoted
$\, \pi(\alpha \vert \, {\mathbf z})$, is
Gamcon~II($\eta^{\prime} / \mu^{\prime}, \, \delta^{\prime}$)
with
\begin{eqnarray}
\delta^{\prime}  =  \delta + k \, ,
\quad
\eta^{\prime} =
\frac{\delta \eta \, + \, \sum_{i = 1}^k z_{i}}
{\delta \, + \, k}
\quad \mbox{and} \quad
\mu^{\prime} =
\mu^{\delta / (\delta \, + \, k)}
\left(\prod_{i = 1}^k z_{i}\right)^{1 / (\delta \, + \, k)} \, ,
\label{eq:posteriorparameters}
\end{eqnarray}
and the conditional density  of $\beta$ given $\alpha$ and ${\mathbf z}$, denoted
$\pi(\beta \vert \, \alpha, \, {\mathbf z})$,
is
Gamma$(\delta^{\prime} \alpha \, + \, 1, \, \delta^{\prime}
\eta^{\prime})$.
\noindent
{\bf \sc Remark~1 . --}
The hyperparameters~$\eta$ and~$\mu$ act on the sufficient statistics
$s_1=\sum_{i = 1}^k z_{i}$ and $s_2=\sum_{i = 1}^k \ln{z_{i}}$, whereas~$\delta$
tunes the importance of these modifications.
When $\delta$ is a positive integer, the introduction of these prior distributions
can loosely be interpreted as artificially adding $\delta$ observations 
with arithmetic mean $\eta$ in $s_1$ and another $\delta$ observations
with geometric mean $\mu$ in $s_2$ with $\eta/\mu=c>1$.
\hfill{$\Box$}

\noi Now the question is: How can we deduce the posterior density of
the GPD parameters $(\alpha, \,\beta)$ given ${\mathbf y}$ (assumed in the following to be a $k-$sample from GPD) from the
gamma conjugate structure, starting with Damsleth prior density
$\pi_{\delta, \, \eta, \mu} (\alpha, \, \beta)$~?

\noi In the remaining of this section the following notations are used:

\noi -- Let  $\theta=(\alpha, \, \beta)$. Consequently $f_{\theta}$ and $g_{\theta}$ will denote respectively p.d.f.'s of GPD($\alpha, \, \beta$) and Gamma($\alpha, \, \beta$) probability distributions.

\noi -- The likelihood function of the observations ${\mathbf y}$ for
$f_{\theta}$ writes $f (  {\mathbf y} \vert \, \theta )$ $=$
$\prod_{i = 1}^k$  $f (  y_i \vert \, \theta )$.
Similarly, the likelihood function of~${\mathbf z}$
for $g_{\theta}$ writes
$g (  {\mathbf z} \vert \, \theta )$
$ = $ $\prod_{i = 1}^k$ $g (  z_i \vert \, \theta )$.

\noi -- We let $\pi(\theta \vert \, {\mathbf y})$ denote the
posterior density of $\theta$ given the observations ${\mathbf y}$
$=$ $(y_1,\dots, y_k)$ corresponding to~$f_{\theta}$ and the prior
density $\pi(\theta)=\pi_{\delta, \, \eta, \mu}$ (Damsleth prior):
\begin{eqnarray}
\pi \left( \theta \left\vert \, {\mathbf y} \right. \right) \, = \,
\frac{f \left(  {\mathbf y} \left\vert \, \theta \right. \right) \,
\pi \left( \theta \right)}{f_{\pi} \left(  {\mathbf y}  \right)} ,
\quad \mbox{where} \;\;\;
f_{\pi}({\mathbf y}) \, = \, \int_{\Theta}
f\left({\mathbf y} \vert \, \theta^{\prime}\right)
\pi(\theta^{\prime}) \, d\theta^{\prime} \, .
\label{eq:posteriorenx}
\end{eqnarray}
\noi -- Similarly, $\pi(\theta \vert \,{\mathbf z})$ denotes
the posterior density of~$\theta$ given
${\mathbf z}$ $=$ $(z_1,\dots, z_k)$
cor\-res\-pon\-ding to $g_{\theta}$ and Damsleth prior density~$\pi$:
\begin{eqnarray}
\pi \left( \theta \left\vert \, {\mathbf z} \right. \right)
\, = \,
\frac{g \left(  {\mathbf z} \left\vert \, \theta \right. \right) \,
\pi \left( \theta \right)}{g_{\pi} \left(  {\mathbf z}  \right)} ,
\quad \mbox{where} \;\;\;
g_{\pi}({\mathbf z}) \, = \, \int_{\Theta}
g \left({\mathbf z} \vert \, \theta^{\prime}\right)
\pi\left(\theta^{\prime}\right) \, d\theta^{\prime} \, .
\label{eq:posteriorenz}
\end{eqnarray}
\noi -- We further denote
\begin{eqnarray}
q_{\pi} \left(  {\mathbf z} \left\vert \, {\mathbf y} \right. \right)
\, = \, \frac{p \left( {\mathbf y} \left\vert \, {\mathbf z} \right. \right)
g_{\pi} \left( {\bf z} \right)}{\int \,
p \left( {\mathbf y} \left\vert \, {\mathbf z}^{\prime}  \right. \right)
g_{\pi} \left( {\mathbf z}^{\prime} \right) \, d{\mathbf z}^{\prime}},
\quad \mbox{where} \quad
p\left({\mathbf y} \left\vert \, {\mathbf z} \right. \right) =
\left(\prod_{i = 1}^k  z_i \right)
\exp{\left( -  \sum_{i=1}^k  y_i z_i \right)}
\, .
\label{eq:densiteintermediaire}
\end{eqnarray}
\noi Thus, the posterior distribution of~$\theta$ given~${\mathbf y}$
is a mixture of the posterior distributions of~$\theta$
given~${\mathbf z}$ with mixing density
$q_{\pi}({\mathbf z} \vert \, {\mathbf y} )$: 
\begin{eqnarray}
\pi \left( \theta \left\vert \, {\bf y} \right.\right)
   \, = \,
   \int
q_{\pi} \left(  {\mathbf z} \left\vert \, {\mathbf y} \right. \right) \,
\pi \left( \theta \left\vert \, {\mathbf z} \right.\right) \,
   d{\mathbf z} \, .
   \label{eq:melangeposteriori}
\end{eqnarray}
It follows that each posterior
moment given ${\mathbf y}$ is the integral of the corresponding posterior
moment given ${\mathbf z}$ with respect to
$q_{\pi} ({\mathbf z} \vert \, {\mathbf y})$.
Unfortunately, the functions $g_{\pi}$
(see (\ref{eq:posteriorenz})--(\ref{eq:densiteintermediaire}))
and $\,{\mathbf z}$ $\mapsto$ $q_{\pi} ({\mathbf z} \vert \,
{\mathbf y})$
(see (\ref{eq:densiteintermediaire})--(\ref{eq:melangeposteriori}))
are not expressible in analytical close form.
Therefore, a numerical algorithm is needed. 
The mixture representation~(\ref{eq:melangeposteriori}) allows us to design a
simple and efficient Gibbs sampler. It is presented in the next
subsection.

\subsection{Gibbs sampling}
\label{Gibbs}
A Gibbs sampler is used to get approximate simulations
from the posterior density of~$\theta$ given ${\mathbf y}$,
$\pi(\theta \vert {\mathbf y})$.
Damsleth's priors are used for $\theta$ $=$ $(\alpha, \, \beta)$.
The proposed sampler generates a Markov chain
whose equilibrium density (denoted $\,\pi({\mathbf z}, \theta \vert \, {\mathbf y})$) is the joint density of
$({\mathbf Z}, \, \theta)$ conditionally on
${\mathbf Y} = {\mathbf y}$ where ${\mathbf Y}$ and ${\mathbf Z}$ denote respectively random vectors of~$k$ independent
Gamma($\alpha$, $\beta$) and GPD($\alpha$, $\beta$) random variables.
To implement the Gibbs sampler we first note that
within the general setting of subsection~\ref{conjugate},
the conditional density
of~$\theta$ given $({\mathbf y}, \, {\mathbf z})$
is independent of ${\mathbf y}$:
$\pi (\theta \vert {\mathbf y}, \, {\mathbf z})$ $=$
$\pi (\theta \vert {\mathbf z})$, and
the conditional density of ${\mathbf Z}$
given $({\mathbf y}, \, \theta)$ is
\begin{eqnarray}
f \left( {\mathbf z}
\left\vert \, {\mathbf y}, \, \theta \right. \right)
\, = \,
\frac{p \left( {\mathbf y}
\left\vert \, {\mathbf z} \right. \right)
g \left(  {\mathbf z} \left\vert \, \theta \right. \right)}
{f \left( {\mathbf y} \left\vert \, \theta \right. \right)}
\,= \, \prod_{i=1}^k
f \left( z_i \left\vert \, y_i, \, \theta \right. \right)
\, \propto \, \prod_{i=1}^k \,
z_i^{\alpha} \, e^{ - \left(\beta \, + \, y_i \right) z_i} \,
\ind_{ z_i > 0 } \, .
\label{eq:densiteconddezsachantx}
\end{eqnarray}
It follows that for $i\leq k$,
conditionally on~$\theta$ and $Y_i = y_i$,
$Z_i$ $\sim$ Gamma$(\alpha + 1, \, \beta + y_i)$\\
independently.
This yields the following intertwining sampler,
where 
$\theta^{(m)}$ denotes the current parameter value
at iteration~$m$. The next iteration:
{\tt
\begin{enumerate}
\item
independently simulates each $z_i^{(m+1)}$ from
Gamma$(\alpha^{(m)} + 1, \, \beta^{(m)} + y_i)$~;
\item
simulates $\theta^{(m + 1)}$ from
$\pi(\theta \vert \, {\mathbf z}^{(m+1)})$.
\end{enumerate}
}
\noindent
In such a setting, both $({\mathbf z}^{(m)})_{m \ge 0}$
and $(\theta^{(m)})_{m \ge 0}$ are Markov chains.
The simulation step of $\theta^{(m + 1)}$ is split into the marginal
simulation of $\alpha^{(m + 1)}$ and the
conditional simulation of $\beta^{(m + 1)}$ given $\alpha^{(m + 1)}$.
Finally, the iteration $m + 1$ of our Gibbs sampler:
{\tt
\begin{enumerate}
\item
independently simulates each $z_i^{(m+1)}$
from Gamma$(\alpha^{(m)} + 1, \, \beta^{(m)} + y_i)$~;
\item
simulates
$\alpha^{(m + 1)}$ from $\pi (\alpha \vert \, {\mathbf z}^{(m+1)})$,
i.e.~from
Gamcon~II($\eta^{\prime} \big/ \mu^{\prime}, \, \delta^{\prime}$)
with $\delta^{\prime}$, $\eta^{\prime}$ and $\mu^{\prime}$
computed from ${\mathbf z}^{(m+1)}$ using
equation~(\ref{eq:posteriorparameters})~;
\item
simulates $\beta^{(m + 1)}$ from
Gamma$(\delta^{\prime} \alpha^{(m + 1)}
\,  +  \, 1, \, \delta^{\prime} \eta^{\prime})$.
\end{enumerate}
}
\noindent When the equilibrium regime is nearly reached, simulated
values of~$\theta$ are approximately issued from the posterior
distribution of~$\theta$ given~${\mathbf y}$.
Implementing the previous algorithm requires
simulating Gamcon~II distributions and choosing adequate values of
the hyperparameters~$\delta$, $\eta$ and~$\mu$ of the priors
$\pi(\alpha)$ and $\pi(\beta \vert \, \alpha)$.
\subsection{Simulating Gamcon~II distributions}
\label{simulation}
Our sampling scheme involves simulations
from Gamcon~II$(c, \,d)$ distributions
with~$c$ $=$ $c^{\prime}$ $=$
$\eta^{\prime} \big/ \mu^{\prime}$,
where $\eta^{\prime}$ and $\mu^{\prime}$
are given by~(\ref{eq:posteriorparameters}),
and moderate to large values
of~$d$ $=$ $d^{\prime}$ $=$ $\delta^{\prime}$ $=$ $\delta + k$.
Up to our knowledge, there is no standard algorithm for
simulating such distributions.  The simulation
method that we present is based on a normal approximation to Gamcon~II
distributions using Laplace's method.

Gamcon~II$(c, \,d)$ can be approximated
by a normal distribution having the same mode.
It is proved in Garrido (2002) that this mode, $M_{c, \, d}$,
is the unique root of the equation
\begin{eqnarray}
\psi\left(d \, M_{c, \, d} \, + \, 1 \right)
\, - \, \psi\left(M_{c, \, d}\right)
\, - \, \ln{d} \, - \, \ln{c} \, = \, 0 \, ,
\label{eq:mode}
\end{eqnarray}
where~$\psi$ denotes the digamma function: the derivative of the logarithm
of~$\Gamma(t)$.
The standard deviation $S_{c, \, d}$ of
the normal approximant distribution
is computed through a Taylor expansion of the Gamcon~II$(c, \, d)$
density in a neighborhood of its mode:
\begin{eqnarray}
S_{c, \, d} \, = \,
\frac{1}{\sqrt{d \psi^{\prime}\left(M_{c, \, d} \right)
\, - \, d^2 \psi^{\prime}\left(d M_{c, \, d} \, + \, 1\right)}}
\, .
\label{eq:variance}
\end{eqnarray}
Garrido (2002) has established that
$(1 - 1/d)/(\ln{c} + \ln{d/2})$ $\le$ $M_{c, \, d}$
$\le$ $2/\ln{c}$.
In practice, $M_{c, \, d}$ is numerically approximated
through the bisection method.

At each iteration, we simulate
Gamcon~II$(c^{\prime}, \, d^{\prime})$ with the help of the
independent Hastings-Metropolis algorithm, which requires
a suitable proposal density. Actually, it is enough
to make only one step of Hastings-Metropolis at each iteration of
the Gibbs sampler: see Robert (1998). 
The proposal density must be as close as possible to the
simulated density, Gamcon~II$(c^{\prime}, \, d^{\prime})$,
and have heavier tails to ensure good mixing.
Since Gamcon~II densities have gamma-like tails,
we cannot directly take the normal approximant
density as a proposal. Rather, we have chosen
a Cauchy proposal density as close as possible to the
normal approximant density to Gamcon~II$(c^{\prime}, \, d^{\prime})$,
i.e.~with the same mode and modal value.
Therefore at each iteration our Hastings-Metropolis step:
{\tt
\begin{enumerate}
\item
independently simulates a new~$Y$ from the Cauchy distribution
with mode\\
$M_{c^{\prime}, \, d^{\prime}}$  and modal value
$1 \big/ (S_{c^{\prime}, \, d^{\prime}} \sqrt{2 \pi})$~;
\item
computes the ratio
$\displaystyle{
\rho \, = \,
\min{\left[1, \, \frac{
f^{\star}_{\mbox{\rm \scriptsize cauchy}}
\left(\alpha^{(m)}\right) \,
\xi_{c^{\prime}, \, d^{\prime}}(Y)}
{f^{\star}_{\mbox{\rm \scriptsize cauchy}} (Y) \;
\xi_{c^{\prime}, \, d^{\prime}}
\left(\alpha^{(m)}\right)} \right]} \, ,}
$
where $f^{\star}_{\mbox{\rm \scriptsize cauchy}}$
denotes the density of $Y$~;
\item
takes $\alpha^{(m+1)}$ $=$ $Y$ with probability~$\rho$ and
$\alpha^{(m+1)}$ $=$ $\alpha^{(m)}$ otherwise.
\end{enumerate}
}

Since all the transition densities involved are positive,
the resulting Markov chain $(\theta^{(m)})_{m \ge 0}$ is ergodic
with unique invariant probability measure equal to the posterior
distribution of~$\theta$ given~${\mathbf y}$.
Intensive numerical experiments reported in Garrido (2002) show
that for $\delta > 0.5$, this Gibbs sampler with one
Hastings-Metropolis step at each iteration converges quickly to its
invariant distribution. Actually, it seems that discarding
the first $500$ iterations is sufficient.
For $\delta \le 0.5$, we observed numerical instabilities.

Bayesian inference on the GPD parameters $\alpha$, $\beta$
is based on the outputs of this algorithm when it has approximately
reached its stationary regime. We then record a sufficient number
of realizations $\alpha^{(m)}$ and $\beta^{(m)}$.

\subsection{Choice of hyperparameters}
\label{hyper}
In this paper we suppose that no expert information is available for the
choice of priors on the GPD parameters $(\alpha, \, \beta)$
and we thus take an empirical Bayes approach.
Introduction of expert information is discussed in 
Appendix~B.  

We take $\delta = 1$ both for convenience (for
$\delta = 1$, the prior $\pi(\alpha)$ reduces to a gamma
distribution, see below) and because a small value of $\delta > 0$
indicates little confidence in prior information (see Remark~1).
Recall that we observed numerical instabilities of our
Gibbs sampler for $\delta < 0.5$.
A natural approach to compute hyperparameter values of priors
$\pi(\alpha)$ and $\pi(\beta \vert \, \alpha)$ is to equate some
of their location parameters ({\it e.g.} the mean) to frequentist
estimates of~$\alpha$ and~$\beta$, denoted here $\widehat{\alpha}$ and
$\widehat{\beta}$. 
Recall that the mean of the prior distribution
$\pi(\beta \vert \, \alpha)$ is $(\alpha + 1) \big/ \eta$. Taking
this mean equal to $\widehat{\beta} = x_{n - k, \, n}$,
the estimate induced by the Hill procedure, and
replacing~$\alpha$ by its Hill estimate
yields
$\eta = (\widehat{\alpha} \, + \, 1) \big/ \widehat{\beta}$.
When $\delta = 1$, the prior $\pi(\alpha)$ reduces to
Gamma$(2, \, \ln{(\eta \big/ \mu)})$. Its
mean is $2 \big/ \ln{(\eta/\mu)}$. Solving
the equation where this prior mean
is set equal to~$\widehat{\alpha}$ yields
$\displaystyle{\mu \, = (\widehat{\alpha} + 1) \;
\left(\exp{\left(- \, 2/\widehat{\alpha}\right)}\right)/\widehat{\beta}}.$

If prior modes are used instead of prior means, a similar approach
leads to slightly different formulas for~$\eta$ and~$\mu$. Actually,
with prior modes explicit expressions can be obtained for
all $\delta > 0$. However, preliminary numerical experiments
yielded better estimates with prior means.

\subsection{Computation of Bayesian estimates}
\label{computation}
When the Gibbs sampler approximately
reaches its stationary regime,
$K$ values denoted $(\alpha^{(m)}, \beta^{(m)})$,
$m  = 1, \dots, K$, are saved.
This sample is used to compute posterior
means, medians or modes to estimate $(\alpha, \, \beta)$,
leading to Bayesian estimates for $(\gamma, \, \sigma)$ and $q_{1-p}$.
\noindent
{\bf \sc Remark~2 . --}
The posterior modes are more difficult to compute
since one has first to construct smooth estimates of the joint
posterior density of~$\alpha$ and~$\beta$.
Numerical experiments reported in Garrido (2002)
for both GPD generated data and excess samples led us to keep only
posterior medians.  \hfill{$\Box$}\\
Concerning the estimation of an extreme quantile $q_{1 - p}$ (${\mathbf y}=(y_1,\dots,y_k)$ is a sample of excesses over a threshold $u$),
a sample of values $\widehat{q}^{(m)}_{1 - p}$ is computed using POT
from the simulated  $(\alpha^{(m)}, \beta^{(m)})$'s:
\begin{equation}
\widehat{q}^{(m)}_{1 - p}
\, = \, u \  + \ \beta^{(m)}
\left[ \left( \frac{np}{k} \right)^{-1/\alpha^{(m)}}
\, - \, 1 \right] \, .
\label{posteriorquantiles}
\end{equation}
Means, histograms and credibility intervals can then be computed and
represented from~(\ref{posteriorquantiles}).
Bayesian credibility intervals for~$\alpha$, $\beta$ and $q_{1-p}$
are obtained by sorting the corresponding simulated values obtained
from the Gibbs sampler.
See sections~\ref{comparative} and~\ref{data} below.
The probability distribution of the observed sample~${\mathbf y}$ can be
estimated  either by GPD$(\widehat{\alpha}, \, \widehat{\beta})$,
or by the posterior predictive distribution.
Similarly, predictive quantile functions can be approximated through
\begin{eqnarray}
\widehat{F^{-1}}_{\mbox{\scriptsize pred}}(y)
\approx \frac{1}{K} \, \sum_{m=1}^K
F^{-1}_{\alpha^{(m)}, \, \beta^{(m)}}(y) \, .
\label{predquantiles}
\end{eqnarray}

\section{Comparative Monte Carlo simulations}
\label{comparative}
Intensive Monte Carlo simulations were used to compare
our Bayes quasi-conjugate estimators (denoted hereafter Bayes-QC) of~$\gamma$ and $q_{1-p}$
with their counterparts
when usual estimators of GPD parameters are used:
Maximum Likelihood (ML), Moment Tail Index estimator (MTI(DEdH)),
Generalized Zipf estimator (ZipfG) and Probability Weighted Moments (PWM).

\subsection{The simulation design}
Three probability distributions in DA(Fr\'echet) were considered
in order to produce various excess samples:

\noi -- The Fr\'echet(1) distribution, for which $\gamma = 1$ and the second-order regular variation parameter (presented below) $\rho = - 1$.
The d.f.~of  Fr\'echet($\beta$) for $\beta > 0$
is $F(x)$ $=$ $\exp{(- x^{- 1/\beta})}$,
$x>0$.

\noi -- The Burr$(1, \, 0.5, \, 2)$ distribution,
for which $\gamma = 1$ and $\rho = - 0.5$.
The d.f.~of Burr($\beta, \, \tau, \, \lambda$)
for $\beta > 0$, $\tau > 0$, $\lambda > 0$ is
$F(x)$ $=$ $1 \, - \, [\beta/(\beta + x^{\tau})]^{\lambda}$,
$x>0$.

\noi -- The Log-Gamma(2) distribution, for which $\gamma = 1$ and $\rho = 0$.
The density of Log-Gamma(2) is
$f(x)$ $=$ $x^{- 2} (\ln{x})^{- 1}$, $x > 0$.

\noi The second-order regular variation
parameter~$\rho$ ($\rho \le 0$) indicates the quality of
approximation of $F_{u}$ by a GPD$(\gamma, \, \sigma(u))$
for high values of~$u$ and suitable $\sigma(u)$'s.
High values of~$\vert \rho \vert$ indicate excellent
fitting, whereas values of~$\vert \rho \vert$ close to~$0$
indicate bad fitting. 

\noi For each one of these three probability distributions,
$100$ data sets of size $n = 500$ were independently generated.
For each simulated data set and each value
of $k$ $=$ $5, 10, \dots, 495$,
we performed $1\,000$ iterations of the Gibbs sampler
and only the last $500$ ones were kept.

\subsection{First results}
For each simulated data set and each value
of $k$ $=$ $5, 10, \dots, 495$ Bayes-QC estimates of~$\gamma$ and $q_{1-p}$, $\, p = 1 / 5\,000$,
are computed as the medians of the resulting
$500$ $\, \widehat{\gamma}^{(m)}$'s and $\, \widehat{q}_{1-p}^{(m)}$'s given in the 500 last iterations of the Gibbs sampler.
Figures~\ref{fig1}--\ref{fig3} display the averages over
the $100$ data sets of these Bayesian estimates
of~$\gamma$ and $q_{1-p}$ as functions of~$k$.
The means and modes were also computed and
gave quite similar results. They are not displayed here.
ML, MTI(DEdH) and ZipfG estimates of~$\gamma$ and  $q_{1-p}$,
$\, p = 1 / 5\,000$, were also computed for each of those
simulated data sets and the same values of~$k$.
Figures~\ref{fig1}--\ref{fig3} display the
averages over the $100$ data sets of these
estimates of~$\gamma$ and  $q_{1-p}$.

The left panels of Figures~\ref{fig1}--\ref{fig3}
show that our estimates of~$\gamma$ based on posterior medians (continuous line curves)
perform rather well compared to the other ones,
and give estimates close to ML and ZipfG.
The right panels of Figures~\ref{fig1}--\ref{fig3} show
that our estimates of $q_{1 - p}$ are very close
to those obtained by ML.
Both give better estimates than MTI(DEdH) but in general
do not perform as well as ZipfG, although they seem to be
more stable as functions of~$k$.
Again, credibility intervals for~$\gamma$ and $q_{1-p}$
and potential improvements when prior information is available
are strong arguments supporting the use of our Bayesian procedure.

\subsection{Bayes credibility and Monte-Carlo confidence intervals}
Monte Carlo simulations were also used to study
whether Bayesian credibility intervals could be used as
approximate frequentist confidence intervals
for~$\gamma$ and $q_{1-p}$.
For each simulated data set, the last $500$ iterations of our Gibbs sampler
provide $500$ estimates $\widehat{\gamma}^{(m)}$ and
$500$ estimates $\widehat{q}_{1-p}^{(m)}$,
$m$ $=$ $501, \dots, 1\,000$.
They can be sorted to provide approximate $90~\%$ credibility intervals
for~$\gamma$ and  $q_{1 - p}$.
The precision of these credibility intervals
was studied through Monte Carlo simulations: for each one of the
three probability distributions considered and for each value of $k$, we counted the number
of simulated data sets (out of $100$) for which the true
values of~$\gamma$ and  $q_{1 - p}$ fell within the
corresponding $90~\%$ credibility intervals.
Figure~\ref{fig4} exhibits the coverage rates
for each simulated distribution
and for $k = 5, 10, \dots, 495$.
These credibility intervals are very accurate for the
Fr\'echet distribution. For the Log-Gamma distribution,
the credibility intervals have good coverage rates
for $q_{1-p}$ but not for~$\gamma$.

This rather unexpected behavior when $\rho = 0$
can be explained in terms of penultimate approximation,
see Diebolt, Guillou and Worms (2003):
it can be proved
that for $\rho = 0$, the distribution of excesses is
better approximated by a GPD with scale parameter
$\gamma + a_{k}(F)$,  where $a_{k}(F)$ is some correction term,
than by a GPD with scale parameter~$\gamma$ (see, e.g.,
the paper coauthored by Kaufmann, pages 183--190
in Reiss and Thomas (2001) along with references
therein and Worms (2001)). This explains why in
the case $\rho = 0$ the estimates of~$\gamma$
strongly deviate from the true value. Furthermore,
Diebolt {\em et al.} (2003) 
have established
for all sufficiently regular estimators
$(\widehat{\gamma}, \, \widehat{\sigma})$ of the parameters
$(\gamma, \, \sigma)$ such as ML or PWM,
that when $\rho = 0$ the estimated survival function
$\bar{F}_{\widehat{\gamma}, \, \widehat{\sigma}}$
is a bias-corrected estimate of
$\bar{F}_{\gamma, \, \sigma}$. We think that this
result partially explains the good and stable coverage rates
observed for $q_{1-p}$ for the Log-Gamma distribution.
These trials show that the credibility intervals
computed through our procedure give very promising
results.

 For each one of the three simulated probability distributions and each value of $k \in \{5,10,\dots,495\}$ the 100 simulated data sets give 100 estimates of $(\gamma,q_{1-p})$. The empirical 0.05 and 0.95 quantiles of the previous estimates give $90\%$ Monte-Carlo Confidence intervals (MCCI) for $(\gamma,q_{1-p})$. The width of these MCCI's are used in Figure \ref{fig6} to compare the precision of our Bayes Quasi-conjugate $q_{1-p}$ estimator to ML, ZipfG and MTI(DEdH) estimators. For Burr data sets (left panel of Figure \ref{fig6}) ZipfG gives the most precise quantile estimators, it is followed by ML, Bayes-QC and MTI(DEdH). For Fr\'echet, Bayes-QC and ML have the best precisions. 
It is worth noting that the Bayes-QC is used here in its empirical version.
Its precision will increase when combined with expert opinions.

\begin{figure}[h]
\epsfxsize=15cm
\epsfysize=10cm
\centerline{
\epsfbox{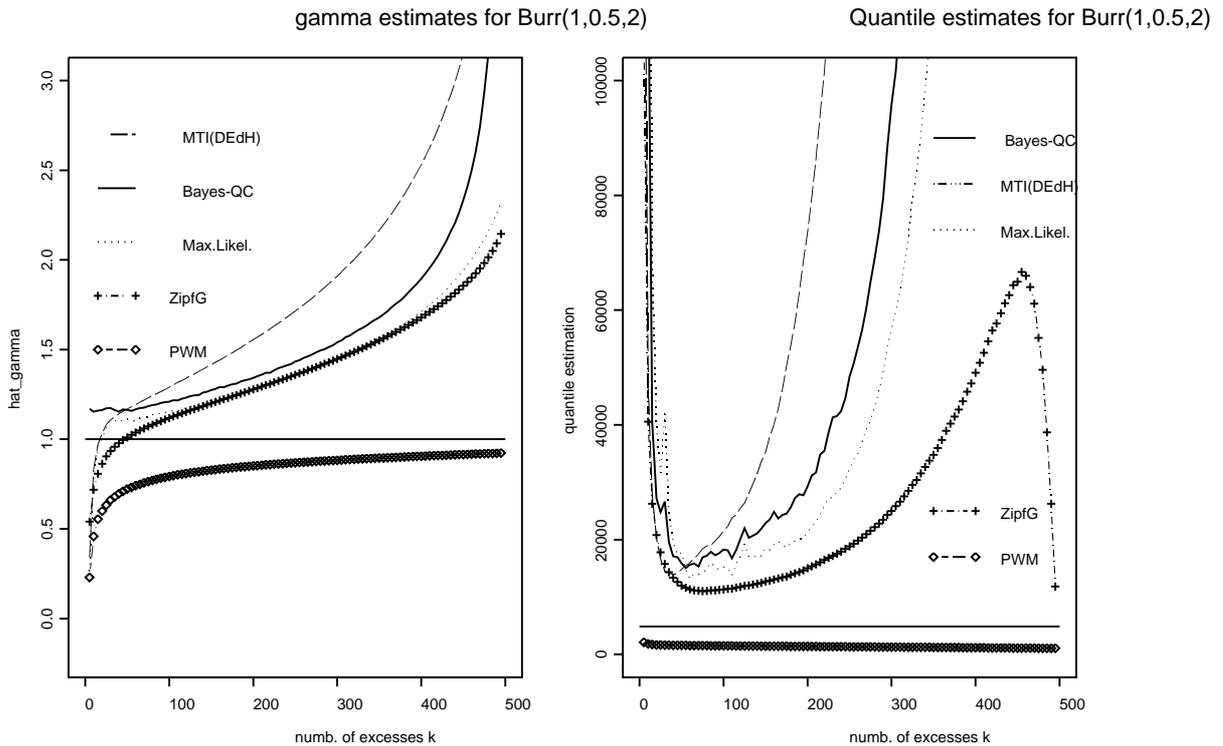}
}
\caption{Means of $\hat{\gamma}$ and $\hat{q}$ on the
$100$ Monte Carlo replications from Burr$(1, \, 0.5, \, 2)$
for different values of the number~$k$ of excesses.}
\label{fig1}
\end{figure}

\begin{figure}[!htbp]
\epsfxsize=15cm
\epsfysize=10cm
\centerline{
\epsfbox{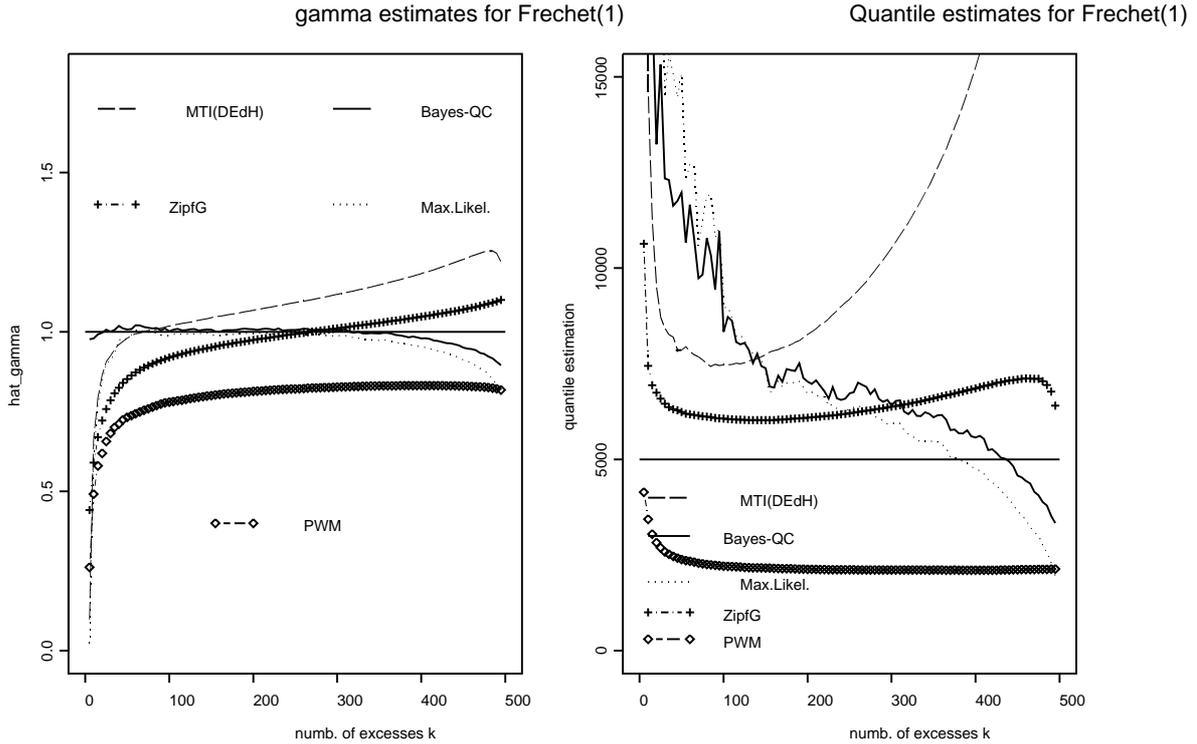}
}
\caption{Means of $\hat{\gamma}$ and $\hat{q}$ on the $100$
Monte Carlo replications from Fr\'echet(1)
for different values of~$k$.}
\label{fig2}
\end{figure}
\begin{figure}[h]
\epsfxsize=15cm
\epsfysize=11cm
\centerline{
\epsfbox{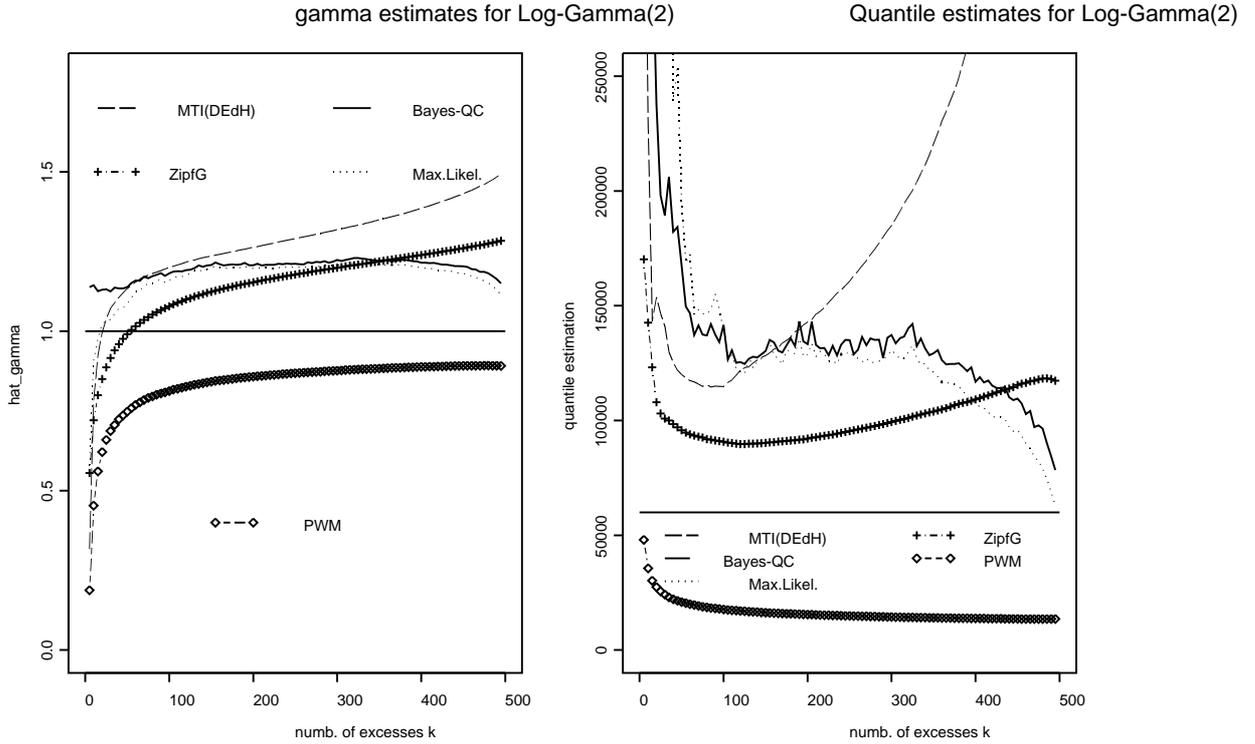}
}
\caption{Means of $\hat{\gamma}$ and $\hat{q}$ on the $100$
Monte Carlo replications from Log-Gamma(2)
for different values of~$k$.}
\label{fig3}
\end{figure}

\begin{figure}[h]
\epsfxsize=15cm
\epsfysize=10cm
\centerline{
\epsfbox{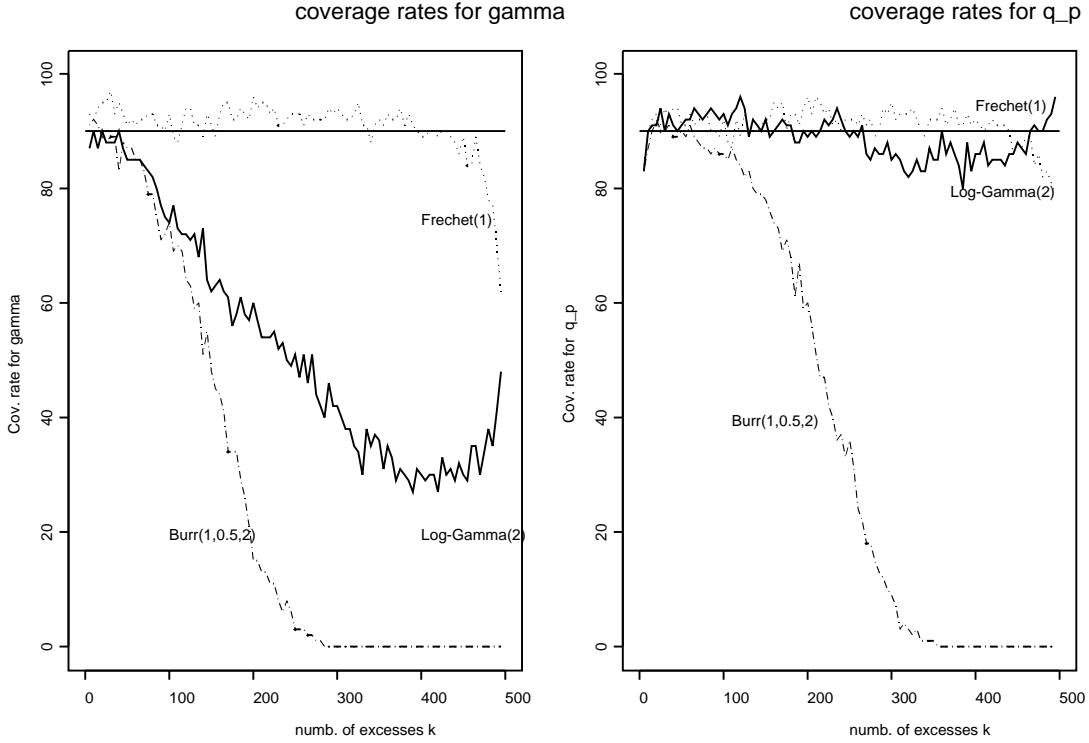}
}
\caption{Coverage rates of our $90~\%$
Bayesian credibility intervals for~$\gamma$ and~$q_{1-p}$
based on the $100$ Monte Carlo replications
for different values of~$k$.}
\label{fig4}
\end{figure}

\begin{figure}[h]
\epsfxsize=15cm
\epsfysize=10cm
\centerline{
\epsfbox{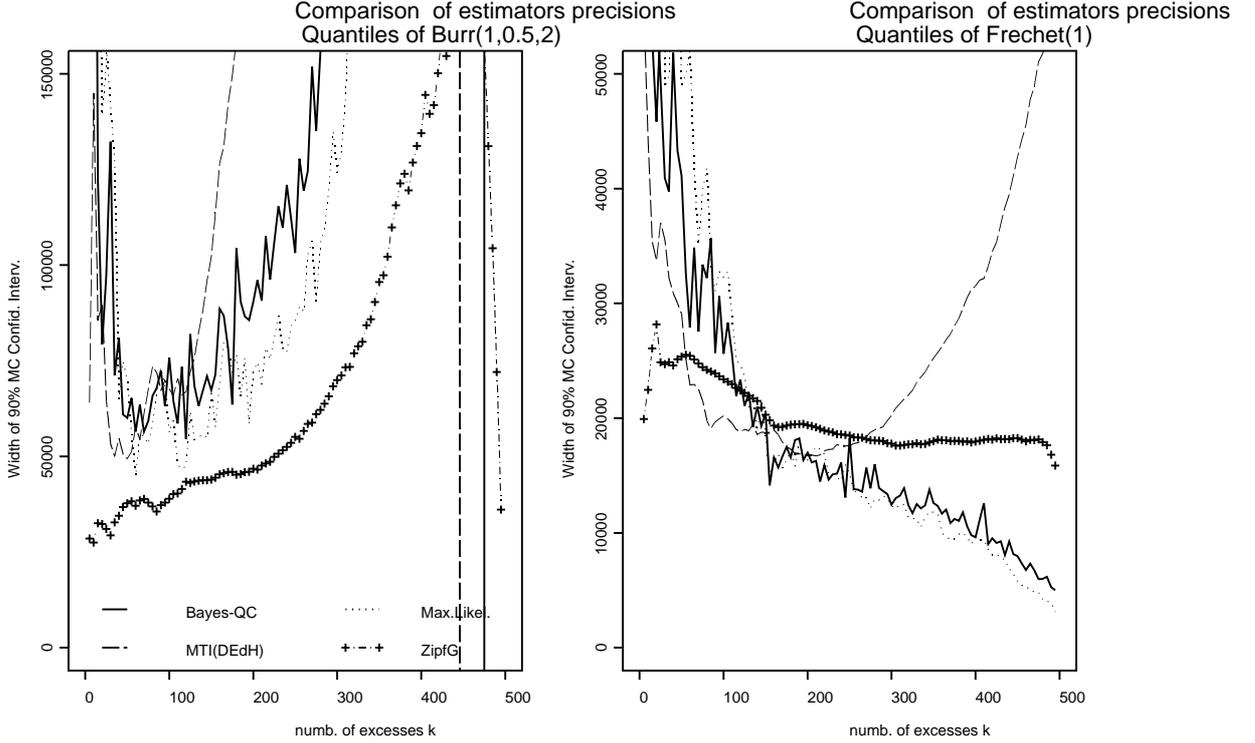}
}
\caption{Width of $90\%$ Monte-Carlo confidence intervals 
of 
~$\hat{q}_{1-p}$ for different values of~$k$.}
\label{fig6}
\end{figure}

\begin{figure}[h]
\epsfxsize=15cm
\epsfysize=10cm
\centerline{
\epsfbox{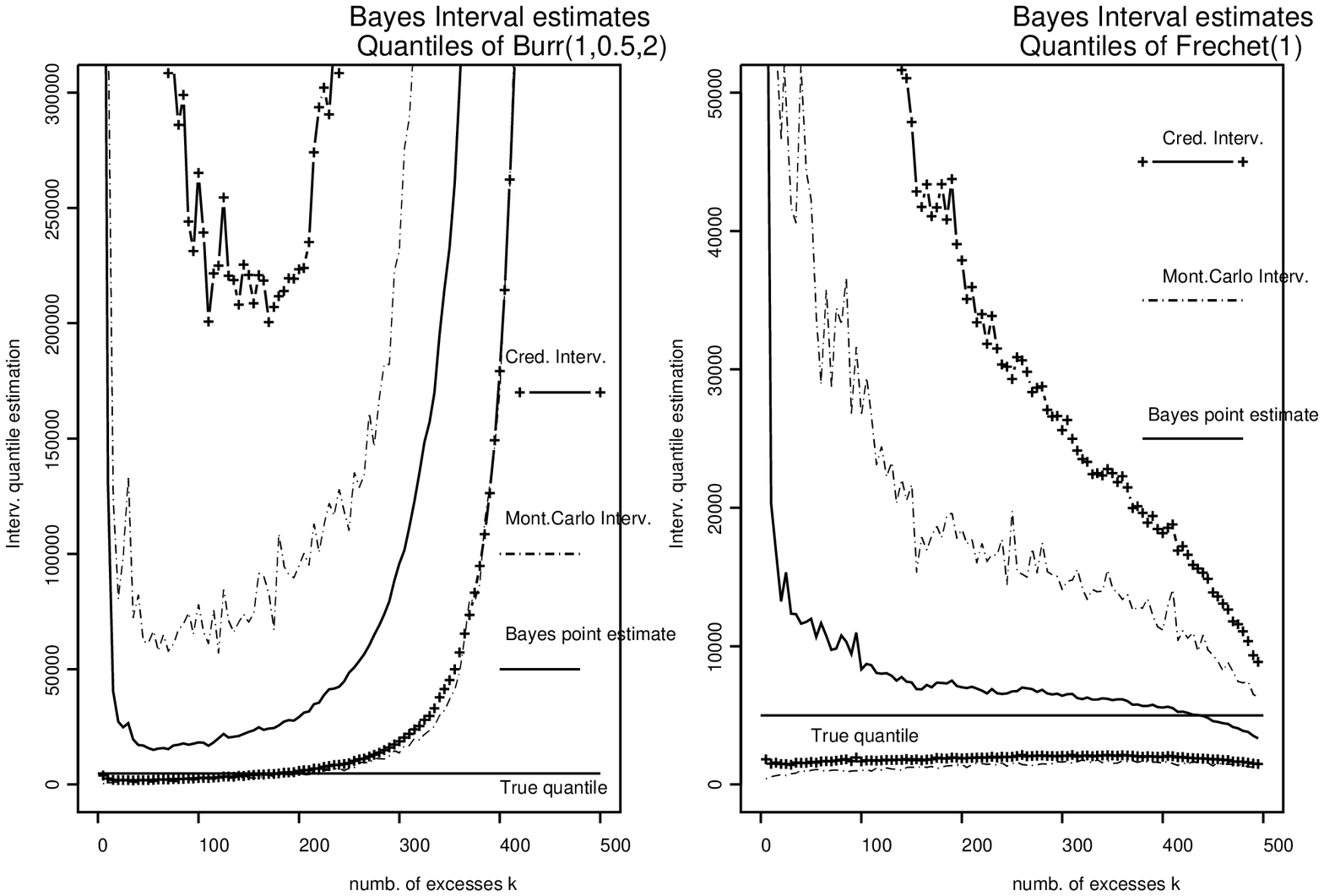}
}
\caption{$90~\%$ Bayes credibility and Monte-Carlo confidence intervals 
for ~$q_{1-p}$ for different values of~$k$.}
\label{fig7}
\end{figure}

In Figure \ref{fig7} the average Bayes credibility intervals (averaged over the 100 simulated data sets) are compared to the $90\%$ Monte-Carlo Confidence intervals (MCCI) for our Bayes-QC estimator. For both Burr (left panel) and Fr\'echet (right panel) ditributions lower
bounds of the average Bayes credibility intervals and MCCI are very close. Upper bounds of average Bayes credibility intervals are larger than those of MCCI. It is interesting to note that the width of the average Bayes credibility intervals are the narrowest for $k$ where Bayes estimate of $q_{1-p}$ is the closest to the true value  $q_{1-p}$. This could be used to chose optimal values of the number of excesses $k$.
Finally, Figure~\ref{fig1}
suggests that the optimal value of~$k$ in the
Burr$(1, \, 0.5, \, 2)$ case is close to~$90$.
For $k = 90$, the credibility intervals
for both~$\gamma$ and $q_{1-p}$ are still satisfactory.

\section{Application to real data sets}
\label{data}
Here, advantages and good performance of our Bayesian estimators
are illustrated through the analysis of extreme events
described by two benchmark real data sets.

Nidd river data are
widely used in extreme value studies
(Hosking and Wallis, 1987
and Davison and Smith, 1990).
The raw data consist in $154$ exceedances of the level~$65$
m$^3$s$^{-1}$ by the river Nidd (Yorkshire, England)
during the period 1934-1969 (35 years).
The $N$-year return level is the water level which is exceeded
on average once in~$N$ years.
Hydrologists need to estimate extreme quantiles
in order to predict return levels over long periods
(250 years, i.e. $p= 9 \, 10^{-4}$, or even 500 years).

Fire reinsurance data were first studied by Schnieper (1993)
and Reiss and Thomas (1999). They represent insurance claims exceeding
$u$ $=$ $22$ millions of Norwegian Kr\"oner from 1983 to 1992 (1985 prices are used).

Many Goodness-of-fit tools (see for example Embrechts {\em et al.} 1997) suggested that excesses from Nidd river data and Fire reinsurance data are well modeled by GPD probability distribution. It was also shown that there are no problems of stationarity violation.

\subsection{Nidd river data}
\begin{figure}[!htbp]
\epsfxsize=15cm
\epsfysize=11cm
\centerline{
\epsfbox{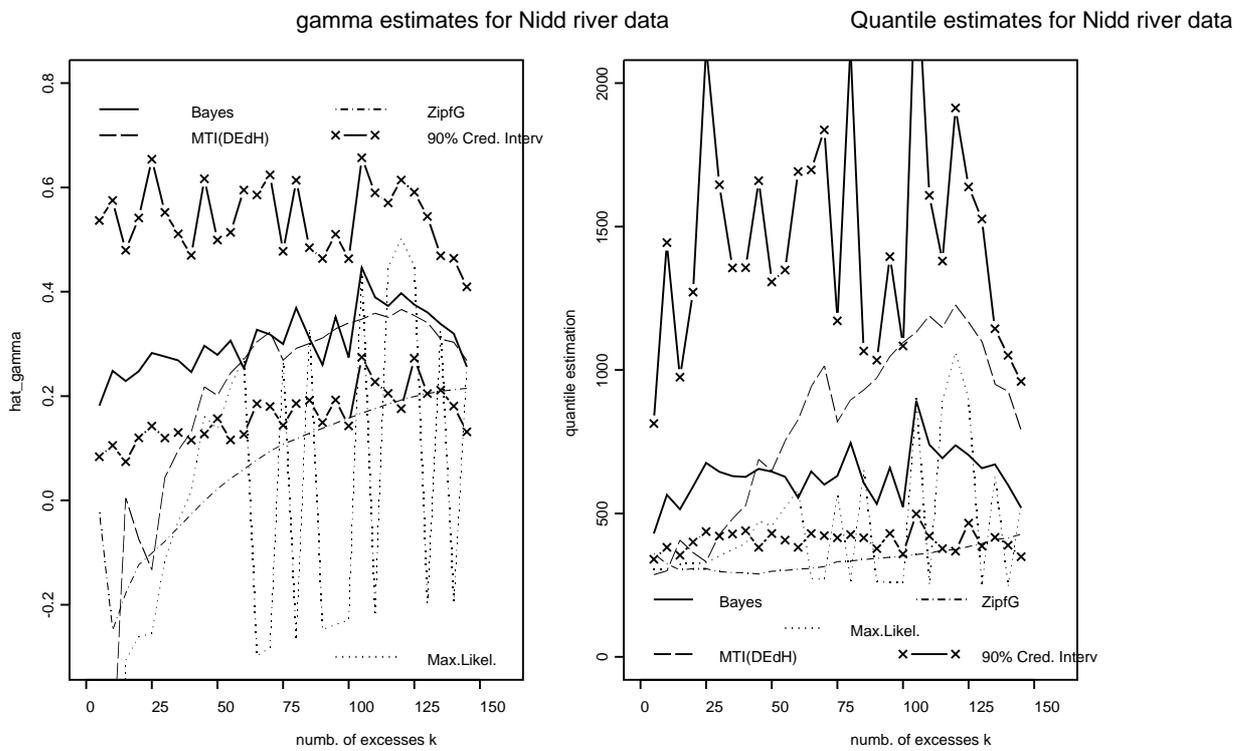}
}
\caption{$\widehat{\gamma}$ and $\widehat{q}_{1-p}$
for the Nidd river data set with several values of~$k$.}
\label{fig5}
\end{figure}

\begin{figure}[!htbp]
\epsfxsize=15cm
\epsfysize=10cm
\centerline{
\epsfbox{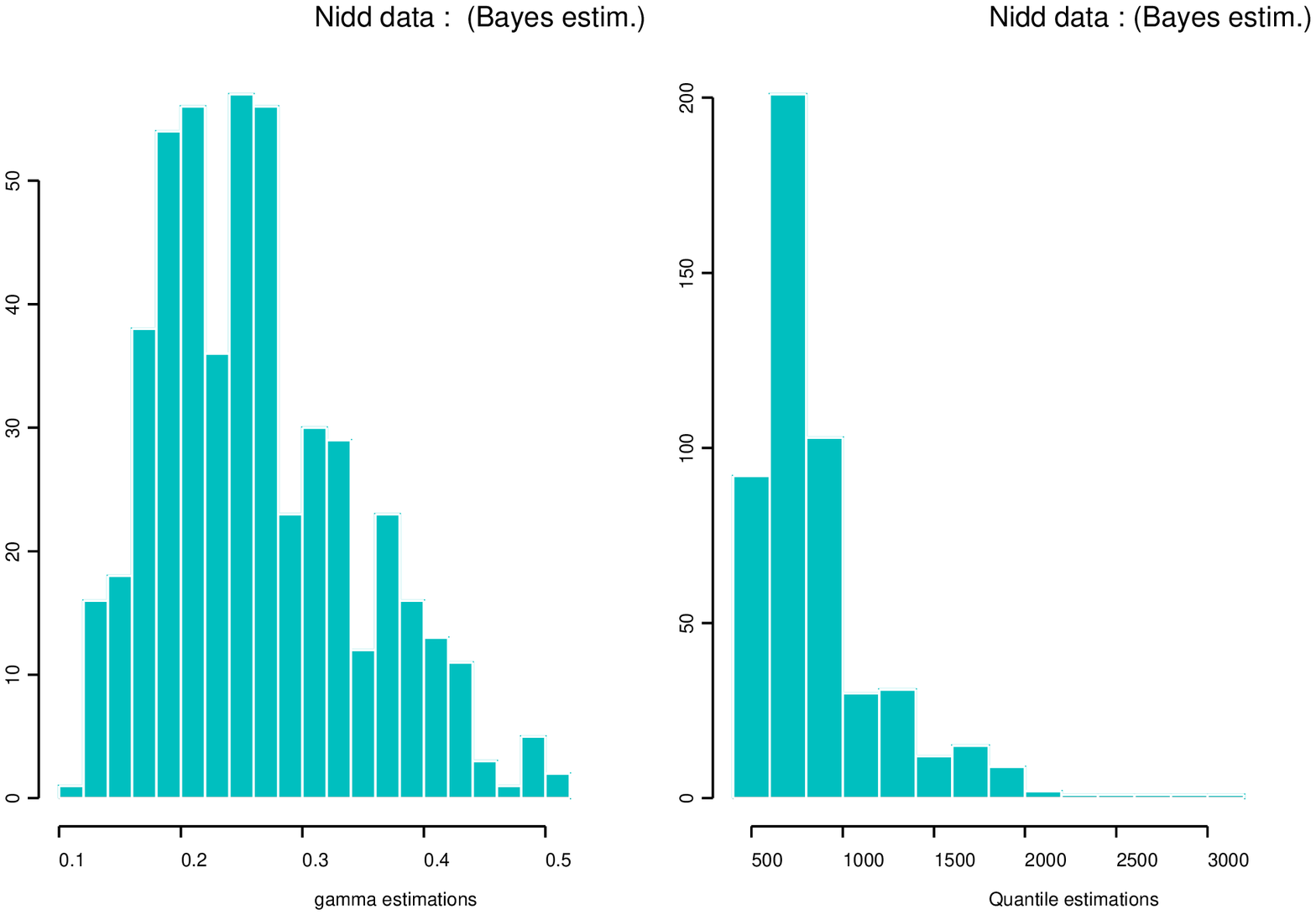}
}
\caption{Histograms of $500$ $\gamma$'s
and $q_{1-p}$'s simulated from the posterior distribution
for the Nidd river data set, $k = 82$.}
\label{fig5b}
\end{figure}

Bayes quasi-conjugate estimates and related $90~\%$ credibility intervals
for~$\gamma$ and $q_{1-p}$ are depicted in Figure~\ref{fig5}
for several values of $k$.
Compared to other estimators, our approach provides the most
stable estimates as $k$ varies.
For~$8$ values of~$k$ Grimshaw's algorithm for computing ML estimates
did not converge: see the broken ML curves in Figure~\ref{fig5}.
Histograms of~$\gamma$'s and $q_{1-p}$'s for $k = 82$
are displayed in Figure~\ref{fig5b}.
Table~\ref{tabres1} summarizes results of the estimation
of the $50$-year and $100$-year return levels of the Nidd river
when the threshold~$u$ is set equal to either $100$ ($k = 39$)
or $120$ ($k = 24$).

\noi
{\bf \sc  Remark~3 . --}
Recall that the $N$-year return level ${RL}_N$
is the water level which is exceeded on average once in~$N$ years.
Equations relating $\qal$ to the return level ${RL}_N$
follow from Davison and Smith (1990):
$\widehat{RL}_N$  $\simeq$
$u   \, -  \, (\widehat{\sigma} \big/ \widehat{\gamma})
[1   -  (\widehat{\lambda} N )^{\widehat{\gamma}}]$,
where it is assumed that the exceedance process is Poisson
with annual rate~$\lambda$.
If we have observed~$k$ excesses above the threshold~$u$
during~$35$ years, then~$\lambda$ is estimated by $k \big/ 35$.
It follows that $\widehat{RL}_N$ $=$ $\widehat{q}_{1 - p}$
with $p$ $\simeq$ $k \, \big/ \, \widehat{\lambda} N n$.
For the Nidd river data, this yields
$p$ $\simeq$ $35 \, \big/ \, N n$.
Therefore, estimating the $100$-year return level is equivalent
to estimating $q_{1 - p}$ by $p$ $=$ $35 \big/ (100 \times 154)$
$\simeq$ $2.27 \, 10^{-3}$.
\hfill{$\Box$}

\noi As shown in Table~\ref{tabres1}, our credibility intervals
with level $95~\%$ (see the last column)
compare well to the Bayesian confidence intervals
obtained by Davison and Smith (1990), which are based on uniform priors
for $q_{1-p}$, $\lambda$ and~$\gamma$
(see Smith and Naylor, 1987 for more details).
Actually, ours are slightly narrower.
\subsection{Fire reinsurance data and net premium estimation}
In the excess-of-loss (XL) reinsurance agreements,
the re-insurer pays only for excesses over
a high value~$u$ of the individual claim sizes.
The {\em net premium} is the expectation of the total claim
amounts that the re-insurer will pay during
the future period $[0, \, T]$:
$\E(S_{N_T})$ $=$ $\sum_{i=1}^{N_{T}} Y_i$,
where $N_T$ is the random number of claims
exceeding~$u$ during $[0, \, T]$ and $Y_1, Y_2, \dots$
are the amounts of excesses above~$u$.
If the $Y_i$'s are modeled by a GPD($\gamma$, $\sigma$)
and the exceedance arrival process is modeled by
a homogeneous Poisson process with annual
rate~$\lambda$, then the net premium over the coming year
is approximated by
$\E(S_{N_1})$ $\simeq$
$\E(N_1) \, \E(Y_i)$ $=$ $\lambda \sigma \big/ (1 - \gamma)$.
Reiss and Thomas (1999, 2001) estimate the net premium
of the Norwegian fire claims reinsurance by analysing
a data set (see Table~\ref{tabdat1})
of large Norwegian fire claims between 1983 and 1992 (1985 prices).
They use a Bayesian inference method for the GPD parameters
and assume that the exceedance process is Poisson.
Independent gamma priors are used for~$\lambda$ and~$\alpha$
and an inverse-gamma prior, with parameters ($a$, $b$), is chosen for~$\beta$.
Posterior means of~$\lambda$, $\alpha$ and~$\beta$ are computed using
Monte Carlo numerical approximations of integrals.
Table~\ref{tabres2} compares estimates of ($\gamma$, $\sigma$)
and the net premium $\lambda \sigma \big/ (1 - \gamma)$
obtained by our quasi-conjugate approach
with those obtained by Reiss and Thomas for ML
and Bayesian estimates with different values of
the hyperparameters~$a$ and~$b$ of the inverse-gamma.
Our approach has the advantage of indicating the precision of
the estimates.
\begin{table}[p]
\begin{center}
\begin{tabular}{c c c c c }
\hline
Analysis &  ML & Bayes-QC & Uniform Bayes & Bayes-QC  \\
& \multicolumn{2}{c}{point Return level estimate} &
\multicolumn{2}{c}{Credibility Intervals} \\
\hline
$50$-year return level &   &  &  &   \\
$u=100$ & 305 & 374 & [210, \, 775] & [266, \, 672] \\
$u=120$ & 280 & 403 & [215, \, 850] & [304, \, 690] \\
$100$-year return level &   &  &  &   \\
$u=100$ & 340 & 457 & [220, \, 935] & [306, \, 911]  \\
$u=120$ & 307 & 499 & [225, \, 940] & [354, \, 961] \\
\hline
\end{tabular}
\caption{Uniform and quasi-conjugate Bayesian
$95~\%$ credibility intervals for $50$-year and $100$-year return levels.
Nidd river data.}\label{tabres1}
\end{center}
\end{table}
\begin{table}[p]
\begin{center}
\begin{tabular}{c c c c }
\hline
year & claim sizes & year & claim sizes  \\
& (in millions) & & (in millions) \\
\hline
1983 &  42.719 & & 23.208 \\
1984 & 105.860 & 1990 & 37.772 \\
1986 & 29.172 & & 34.126 \\
& 22.654 & & 27.990 \\
1987 &61.992 & 1992 & 53.472 \\
& 35.000 && 36.269 \\
1988 & 26.891 & & 31.088 \\
1989 & 25.590 & & 25.907 \\
& 24.130 & & \\
\hline
\end{tabular}
\caption{Norwegian fire claims sizes over
22.0 millions NKr from 1983 to 1992 (1985 prices),
from Schnieper (1993).}\label{tabdat1}
\end{center}
\end{table}
\begin{table}[p]
\begin{center}
\begin{tabular}{c c c c c }
\hline
Analysis &  $\widehat{\gamma}$ & $\widehat{\sigma}$ &
\multicolumn{2}{c}{Net premium}  \\
& & & Point estimate  & $90~\%$ credib.~interv. \\
\hline
ML for GPD  & 0.254 & 11.948 & 27.23 & \\
\\
Bayes (Reiss and Thomas) & & & &\\
Inv.-Gamma($a=2$, $b=2$) & 0.288 & 11.658 &  27.83 & \\
Inv.-Gamma($a=4$, $b=6$) & 0.274 & 11.814 &  27.66 & \\
\\
Bayes-QC approach & 0.384 & 10.332 & 30.03 & [17.09, \, 84.39] \\
\hline
\end{tabular}
\caption{Bayesian estimates of $\gamma$, $\sigma$
and the XL net premium for fire reinsurance data.}
\label{tabres2}
\end{center}
\end{table}

\section{Discussion}
The proposed quasi-conjugate Bayes approach has many advantages when compared
to standard GPD parameters and extreme quantiles estimators:

\noi -- it can incorporate experts prior information and improve estimation of extreme events even when data are scarce,

\noi -- it provides Bayes credibility intervals assessing the quality of the extreme events estimation,

\noi -- it often gives estimators with weak dependence on the number $k$ of used excesses,

\noi -- the variances of the empirical quasi-conjugate Bayes estimators compares well to the variances of the standard estimators. This suggests that quasi-conjugate Bayes estimators including experts opinion would give very accurate extreme quantile estimators, this point will be illustrated in a forthcoming paper.

We deeply describe the proposed quasi-conjugate Bayes approach for the most frequent case of DA(Fr\'echet) where tails are heavy ($\gamma>0$), 
the case of DA(Gumbel) is analytically discussed in Appendix~A. Future work is needed to extend this approach to the general case where the user has no prior idea on $\gamma$.

The present paper is the first of a series of
papers devoted to various developments of the Bayesian inference
methodology that we introduced here. In particular, we will
study
how to determine and compute hyperparameters in a hierarchical
structure setting based on the quasi-conjugate class defined here
to take into account realistic expert prior information on extreme events.

Finally, note that it would be possible to include a Poisson
parameter for the stream of exceedances as in Reiss and Thomas
(2000). Also, spatial quantile estimation and multivariate or time-series
extensions based on our approach are natural and very promising.
\begin{center}
{\bf Appendix A}
\end{center}

We present here a brief account of the simple case
where
Bayesian inference is made for exponential distributions,
rather than GPD's
with both parameters unknown. This simple setup
is of interest since it is
the Bayesian analogue of the Exponential
Tail (ET) method  (Breiman {\em et al.} 1990), and all
calculations lead to explicit analytical formulas.

Set $\lambda = 1/\sigma$ and
$f_{{\lambda}}(y)$ $=$ $\lambda\,e^{- \lambda y}\,\mathbf{1}_{y > 0}$.
Denote by $\pi_{a, \, b}$ the prior Gamma$(a, \, b)$ density
($a, \, b$ $> 0$).
The posterior density
$\pi_{a, \, b}(\lambda \vert \, y_{1}, \dots, y_{k})$ is
Gamma$(a + k, \, b + S_{k})$, where $S_{k}$ $=$ $\sum_{i=1}^k y_{j}$.
Expert information is reflected in the choice
of~$a$ and~$b$. The corresponding posterior predictive distribution
is GPD $(a + k, \, b + S_{k})$, with
$\gamma_{\mbox{\rm \scriptsize pred}}$ $=$ $1 / (a + k)$ and
$\sigma_{\mbox{\rm \scriptsize pred}}$ $=$ $(b + S_{k}) / (a + k)$.
Our first estimate of $q_{1 - p}$ is based on the posterior mean
$\widehat{\lambda}_{\mbox{\rm \scriptsize bayes}}$ $=$
$(a + k) \big/ (b + S_{k})$ of $\lambda$:
$$
\widehat{q}_{1- p, \; \mbox{\rm \scriptsize bayes}} \; = \;
u \; + \; \frac{b + S_{k}}{a + k}
\ln{\left( \frac{k}{np} \right)} \, .
$$
Our second estimate is based on the posterior predictive
distribution:
$$
\widehat{q}_{1 - p, \; \mbox{\rm \scriptsize pred}} \; = \;
u \; + \; \left( b + S_{k} \right)
\left[\left( \frac{k}{np} \right)^{1 / (a + k)}
\; - \; 1 \right] \, .
$$
Our third estimate is based on the transformed posterior
distribution. Since the posterior on
$\sigma$ $=$ $1 / \lambda$ is an inverse-gamma distribution
with density
$$
\frac{\left( b + S_{k} \right)^{a + k}}
{\Gamma \left(a + k\right) \, \sigma^{a + k + 1}}  \,
\exp{\left(- \, \frac{b + S_{k}}{\sigma} \right)} \,
\mathbf{1}_{\sigma > 0}
$$
the corresponding distribution of
$u  + \sigma \ln{( k / np )}$ has a similar shape,
and has mean
$$
\widehat{q}_{1 - p, \; \mbox{\rm \scriptsize post}}
\; = \; u \; + \;  \frac{b + S_{k}}{a + k - 1} \,
\ln{\left(\frac{k}{np}\right)}
$$
For~$k$ large enough,
$\widehat{q}_{1 - p, \; \mbox{\rm \scriptsize bayes}}$
is close to
$\widehat{q}_{1 - p, \; \mbox{\rm \scriptsize post}}$
with respect to the standard deviation scale, which is
of the order of
$(b + S_{k}) (a + k)^{- 3/2} \ln{( k / np )}$.
On the contrary, a Taylor expansion shows that
when $\ln{( k / np )} / (a + k)$ is not too large,
$$
\widehat{q}_{1 - p, \; \mbox{\rm \scriptsize pred}} \;
\approx \;
u \; + \;
\frac{b + S_{k}}{a + k} \ln{\left( \frac{k}{np}\right)}
\left[1 \; + \;
\frac{\ln{\left( \frac{k}{np} \right)}}{2(a + k)}\right] \, .
$$
The distance between
$\widehat{q}_{1 - p, \; \mbox{\rm \scriptsize pred}}$
and each of the two other estimates can be significant,
and $\widehat{q}_{1 - p, \; \mbox{\rm \scriptsize pred}}$
exhibits a positive bias with respect to the other estimates.
We have observed a similar behavior when dealing with GPD's:
This is the reason why we have discarded the analogous of
$\widehat{q}_{1 - p, \; \mbox{\rm \scriptsize pred}}$ in
that setting, and selected estimates of $q_{1 - p}$
based on its posterior distribution.

\begin{center}
{\bf Appendix B}
\end{center}

We propose here two examples for introducing
expert opinion in our Bayesian framework.
In the first one, we use a partial expert opinion:
It acts only on one parameter of the GPD distribution,
whereas the second one is derived from the empirical
choice made in Subsection~\ref{hyper}. 
In the second one, the expert opinion acts on both
parameters of the GPD distribution.


\paragraph{Example 1.}

In this situation, the expert 
provides a rare value $q_{\mbox{\scriptsize max}}$ of the variable
as well as an interval $[p_1,\,p_2]$ containing the probability  $p$
to overpass $q_{\mbox{\scriptsize max}}$ and a (small) probability 
$\varepsilon$ measuring the uncertainty of this opinion.
From Pickands theorem, we deduce
$ \tilde q_{1-p_2} \le q_{\mbox{\scriptsize max}} \le \tilde q_{1-p_1}$
 where
\begin{equation} 
\label{estiquan}
\tilde q_{1-p} = u + \beta \left[ \left(
 \frac{np}{k} \right)^{-1 / \alpha} -1 \right].
\end{equation} 
Plugging $u=x_{n-k,n}$ yields
\begin{equation} \label{traduc-avis-expert}
\beta \left[ \left( \frac{np_2}{k} \right)^{-1/\alpha} -1 \right] \le
           q_{\mbox{\scriptsize max}} - x_{n-k,n} \le
    \beta \left[ \left( \frac{np_1}{k} \right)^{-1/\alpha} -1 \right] .
\end{equation}
Replacing $\alpha$ by its Hill estimate $\hat\alpha$
(similarly to Subsection~\ref{hyper}),
we obtain the following bounds for $\beta$:
$$ \beta_1 \le \beta \le \beta_2 \ \ \textrm{ where } \ \ \beta_1 =
   \displaystyle  \frac{q_{\mbox{\scriptsize max}} - x_{n-k,n}}{
 (n p_1/k)^{-1/\hat{\alpha}} - 1} \ \ \textrm{ and } \ \ \beta_2 = \frac{
q_{\mbox{\scriptsize max}} - x_{n-k,n}}{(np_2/k)^{-1/\hat{\alpha}} -1}\,.$$
Note that, the converse approach (fixing $\beta$ and bounding $\alpha$)
can also be considered.
The prior distribution of $\beta$ given $\alpha = \hat{\alpha}$
is Gamma$(\delta \hat{\alpha} + 1,\, \delta \eta)$.
Suppose that $[0,\beta_1]$ and $[\beta_2,+\infty)$ both have probability
$\varepsilon/2$ for this gamma distribution. We thus have two
nonlinear equations permitting to obtain $\delta$ and $\eta$. 
Approaching the gamma distribution by a Gaussian one yields
explicit solutions:
\begin{eqnarray}
\displaystyle  \delta & = & \frac{1}{\hat{\alpha}} \left[ z_{1-\varepsilon/2}^2
   \left( \frac{\beta_1 + \beta_2}{\beta_2 - \beta_1} \right)^2 - 1 \right] .\\
\displaystyle  \eta & = &  2 \hat{\alpha} z_{1-\varepsilon/2} \, \frac{\beta_1
   + \beta_2 }{ (\beta_2 - \beta_1)^2 }  \left[  z_{1-\varepsilon/2}^2
 \left( \frac{\beta_1 + \beta_2}{\beta_2 - \beta_1} \right)^2 - 1 \right]^{-1},
\end{eqnarray}
where $z_{1-\varepsilon/2}$ is the $1-\varepsilon/2$ quantile of the
standard Gaussian distribution.
The parameter $\mu$ is determined by imposing that the
gamconII($\delta,\,\eta/\mu$) prior distribution of $\alpha$
has mode $\hat{\alpha}$. From~(\ref{eq:mode}), we obtain:
\begin{equation}
\displaystyle  \mu = \eta \exp \left[ \ln \delta + \psi (\hat{\alpha} ) - \psi
\left( \delta\hat \alpha +1 \right) \right] ,
\end{equation}
where $\psi$ denotes the digamma function.

\paragraph{Example 2.}

Here, the expert 
provides two rare values $q_{\mbox{\scriptsize max},1}$ and
$q_{\mbox{\scriptsize max},2}$ of the variable
as well as their associated probabilities $p_1$ and $p_2$
to be overcome. The confidence on this opinion is
measured by $\delta$ (see Remark~1).
Equation~(\ref{estiquan}) yields two nonlinear equations
from which the GPD parameters $(\alpha_0,\beta_0)$ can
be computed.
This allows to determinate the hyperparameters $\eta$ and $\mu$.
To this end, we impose $\alpha_0$ and $\beta_0$ to be respectively
the modes of the prior distribution of $\alpha$ and $\beta$ given $\alpha$.
From~(\ref{eq:mode}), we obtain
\begin{eqnarray}
\eta&=&\alpha_0/\beta_0 \\
\mu&=&\eta \exp(\ln \delta + \psi(\alpha_0) -\psi(\alpha_0\delta +1)),
\end{eqnarray}
where $\psi$ denotes the digamma function.



\end{document}